%Paper: gr-qc/9509004
%From: ESPOSITO@napoli.infn.it
%Date: Sun, 3 Sep 1995 20:20:42 +0200 (CET-DST)

\magnification \magstep1
\raggedbottom
\openup 2\jot
\voffset6truemm
\headline={\ifnum\pageno=1\hfill\else
{\it THE SINGULARITY PROBLEM FOR SPACE-TIMES
WITH TORSION} \hfill \fi}
\centerline {\bf THE SINGULARITY PROBLEM FOR}
\centerline {\bf SPACE-TIMES WITH TORSION}
\vskip 1cm
\centerline {GIAMPIERO ESPOSITO}
\vskip 1cm
\noindent
{\it Department of Applied Mathematics and Theoretical Physics,
Silver Street, Cambridge CB3 9EW, U.K.}
\vskip 0.3cm
\noindent
{\it St. John's College, Cambridge CB2 1TP, U.K.}
\vskip 1cm
\noindent
{\bf Summary.} - The problem of a rigorous theory of
singularities in space-times with torsion is addressed.
We define geodesics as curves whose tangent vector moves by
parallel transport. This is different from what other authors
have done, because their definition of geodesics only involves
the Christoffel connection, though studying theories with
torsion. We propose a preliminary definition of singularities
which is based on timelike or null geodesic incompleteness,
even though for theories with torsion the paths of particles
are not geodesics. The study of the geodesic equation for
cosmological models with torsion shows that the definition
has a physical relevance. It can also be motivated, as
done in the literature, remarking that the causal structure
of a space-time with torsion does not get changed with
respect to general relativity. We then prove how to extend
Hawking's singularity theorem without causality assumptions
to the space-time of the ECSK theory. This is achieved
studying the generalized Raychaudhuri equation in the ECSK
theory, the conditions for the existence of conjugate points
and properties of maximal timelike geodesics. Hawking's
theorem can be generalized, provided the torsion tensor obeys
some conditions. Thus our result can also be interpreted as
a no-singularity theorem if these additional conditions
are not satisfied. In other words, it turns out that the
occurrence of singularities in closed cosmological models
based on the ECSK theory is less generic than in general
relativity. Our work is to be compared with previous papers in
the literature. There are some relevant differences, because
we rely on a different definition of geodesics, we keep the
field equations of the ECSK theory in their original form
rather than casting them in a form similar to general
relativity with a modified energy momentum tensor, and we
emphasize the role played by the full extrinsic curvature
tensor, which now contains torsion.
\vskip 1cm
\leftline {\bf 1. - Introduction.}
\vskip 1cm
The singularity problem is still of vital importance in
modern cosmology [1,2] and the aim of this paper is the
study of singularities for classical theories of
gravitation. Indeed, the singularity theorems of Penrose,
Hawking and Geroch [3-7] show that Einstein's general
relativity leads to the occurrence of singularities in
cosmology in a rather generic way. On the other hand, much
work has also been done on alternative theories of gravitation.
In particular, it is by now well known that, when we describe
gravity as the gauge theory of the Poincar\'e group, this
naturally leads to theories with torsion [8,9]. Thus it is
natural to address the question: is there a rigorous theory
of singularities in a space-time with torsion ? To this
purpose, at first we present in sect. {\bf 2} and {\bf 3} a
brief review of the space-time manifold and of the
definition of singularities in general relativity. We then
define in sect. {\bf 4} geodesics in space-times with
torsion, and we discuss their relevance for cosmology.
In sect. {\bf 5}, we prove how to extend Hawking's
singularity theorem without causality assumptions to the
space-time of the Einstein-Cartan-Sciama-Kibble
(hereafter referred to as ECSK) theory. A comparison with
previous important work appeared in [10] is also made.
Finally, our approach and our results are summarized in
sect. {\bf 6}.
\vskip 1cm
\leftline {\bf 2. - The space-time manifold.}
\vskip 1cm
A space-time $(M,g)$ is the following collection of mathematical entities
[1,11]:

1) A connected four-dimensional Hausdorff $C^{\infty}$ manifold $M$ ;

2) A Lorentz metric $g$ on $M$, namely the assignment of a nondegenerate
bilinear form $g_{\mid p}:T_{p}M$x$T_{p}M\rightarrow R$ with diagonal form
$(-,+,+,+)$ to each tangent space. Thus $g$ has signature $+2$ and is not
positive-definite;

3) A time orientation, given by a globally defined timelike vector field
$X : M\rightarrow TM$. A timelike or null tangent vector $v\in T_{p}M$ is
said to be future (respectively, past)
directed if $g(X(p),v)<0$,
(respectively, $g(X(p),v)>0$).

Some important remarks are now in order:

a) Condition (1) can be formulated for each number of space-time
dimensions $\geq 2$.

b) Also the convention $(+,-,-,-)$ for the diagonal form of the metric
can be chosen. This convention seems to be more useful in
the study of spinors, and can also be adopted in using tensors as Penrose
does so as to avoid a change of conventions. The definitions of timelike
and spacelike will then become:
$X$ is timelike if $g(X(p),X(p))>0$, $\forall p \in M$, and $X$ is spacelike
if $g(X(p),X(p))<0$ $\forall p \in M$.

c) The pair $(M,g)$ is only defined up to equivalence. Two pairs $(M,g)$
and $(M',g')$ are equivalent if there is a diffeomorphism
$\alpha: M\rightarrow M'$ such that: $\alpha_{*}g=g'$. Thus we are
really dealing with an equivalence class of pairs [1].

A concept which will be very useful in sect. {\bf 5} is
the one of Lorentzian arc length.
Let $\Omega_{pq}$ be the space of all future directed nonspacelike curves
$\gamma: [0,1]\rightarrow M$ with $\gamma(0)=p$ and $\gamma(1)=q$. Given
$\gamma \in \Omega_{pq}$ we choose a partition of $[0,1]$ such that $\gamma$
restricted to $[t_{i},t_{i+1}]$ is smooth $\forall i=0,1,...,n-1$. The
Lorentzian arc length is then defined as [11]
$$
L(\gamma)\equiv \sum_{i=0}^{n-1}\int_{t_{i}}^{t_{i+1}}
\sqrt{-g(\gamma'(t),\gamma'(t))} \; dt \; .
\eqno (2.1)
$$
So as to avoid confusion in comparing with the convention used
in [2,8-10], we wish to emphasize that we use the term
Riemannian geometry for the case of positive-definite metrics
(see [12,13]), whereas general relativity is more properly
called a Lorentzian theory [1, 11, 14].
\vskip 10cm
\leftline {\bf 3. - The definition of singularities in
general relativity.}
\vskip 1cm
The singularity theorems in general relativity [1]
were proved using a definition of singularities based on the $g$-boundary.
Namely, one defines a topological space, the $g$-boundary, whose points
are equivalence classes of incomplete nonspacelike geodesics. The points
of the $g$-boundary are then the singular points of space-time. As
emphasized for example in [15], this definition has two basic
drawbacks:

1) it is based on geodesics, whereas in [16] it was proved there
are geodesically complete space-times with curves of finite length and
bounded acceleration;

2) there are several alternative ways of forming equivalence classes
and defining the topology.

Schmidt's method is along the following lines [15].
Connections are known to provide a parallelization of the
bundle $L(M)$ of linear frames.
This parallelization can be used to define a Riemannian metric,
which has the effect of making a connected component of
$L(M)$ into a metric space. This connected component $L'(M)$ is dense in
a complete metric space $L_{C}'(M)$.
One defines ${\overline M}$ as the set of orbits of the
transformation group on $L_{C}'(M)$, and
the $b$-boundary $\partial M$ of $M$ is then defined as:
$\partial M \equiv {\overline M}-M$. Finally,
singularities of $M$ are defined as points of the $b$-boundary
$\partial M$ which are contained in the $b$-boundary of any extension of
$M$.

However, also Schmidt's definition has some drawbacks [17].
In fact in a closed Friedmann-Robertson-Walker (hereafter referred
to as FRW) universe the initial and final singularities form the
same single point of the $b$-boundary [18], and
in the FRW and Schwarzschild solutions the $b$-boundary points are
not Hausdorff separated from the corresponding space-time [19].
A fully satisfactory improvement of Schmidt's definition is still an open
problem. Unfortunately, a recent attempt appeared in [17] was not correct.
In the next sections we will not use this formal apparatus. But we
thought it was worth summarizing this kind of mathematical
results in a paper devoted to the study of the singularity
problem.
\vskip 1cm
\leftline {\bf 4. - Space-times with torsion and their geodesics.}
\vskip 1cm
A space-time with torsion (hereafter referred to as $U_{4}$ space-time) is
defined adding the following fourth requirement to the ones in
sect. {\bf 2}:

4) Given a linear $C^{r}$ connection ${\widetilde \nabla}$ which obeys
the metricity condition, a nonvanishing tensor
$$
S(X,Y)={\widetilde \nabla}_{X}Y -{\widetilde \nabla}_{Y}X -[X,Y]
\; ,
\eqno (4.1)
$$
where $X$ and $Y$ are arbitrary $C^r$ vector fields and the square bracket
denotes their Lie bracket. The tensor ${S\over 2}$
is then called the torsion tensor (compare with [1]).

Now, it is well known that the curve $\gamma$ is defined to be a
geodesic curve if its
tangent vector moves by parallel transport, so that $\nabla_{X}X$
is parallel to ${\left({\partial \over \partial t}\right)}_{\gamma}$
(see, however, comment before our definition of singularities).
A new parameter $s(t)$, called affine parameter, can always be found such
that, in local coordinates, this condition is finally expressed by the
equation
$$
{d^{2}x^{a}\over ds^{2}}+\Gamma_{bc}^{\; \; \; a}
{dx^{b}\over ds}{dx^{c}\over ds}=0 \; .
\eqno (4.2)
$$
The geodesic equation (4.2) will now contain the effect of torsion through
the symmetric part of the connection coefficients:
$$
\Gamma_{(bc)}^{\; \; \; \; \; a}
= \left\{{a \atop bc}\right\}
+2S_{\; \; (bc)}^{a} \; ,
$$
where $S_{\; \; (bc)}^{a}$ is
not to be confused with the vanishing $S_{(bc)}^{\; \; \; \; \; a}$.
It is very useful to study this equation
in a case of cosmological interest. For example, in a closed FRW universe
the only nonvanishing components of the torsion tensor are the ones given
in [20]
$$
S_{m0}^{\; \; \; m}=Q(t) \; , \; \; \; \;
\forall m=1,2,3,
$$
so that (4.2) yields
$$
{d^{2}x^{0}\over ds^{2}}+a{da\over ds}{ds\over dt}c_{ii}
{\left({dx^{i}\over ds}\right)}^{2}=0 \; ,
\eqno (4.3)
$$
$$
{d^{2}x^{m}\over ds^{2}}+\Gamma_{ij}^{\; \; \; m}{dx^{i}\over ds}
{dx^{j}\over ds}
+2\left({1\over a}{da\over ds}{ds\over dt}-Q\right)
{dx^{0}\over ds}{dx^{m}\over ds}=0 \; .
\eqno (4.4)
$$
In (4.3), $c_{ii}$ are the diagonal components of the unit three-sphere
metric, and we are summing over all $i=1,2,3$. In (4.4), we used the
result of [20] according to which
$$
\Gamma_{0m}^{\; \; \; \; m}={{\dot a}\over a}-2Q
\; , \; \; \; \;
\Gamma_{m0}^{\; \; \; \; m}={{\dot a}\over a}
\; , \; \; \; \;
\forall m=1,2,3  \; .
\eqno (4.5)
$$
Of course, $\dot a$ denotes ${da\over ds}{ds\over dt}$.
One can use the relation $c_{ij}={\rm diag}
\Bigr(1,(\sin \chi)^{2},(\sin \chi)^{2}(\sin \theta)^{2}
\Bigr)$ to compute the connection coefficients
$\Gamma_{ij}^{\; \; \; m}$ and write down a more explicit form
of (4.3), (4.4). However, we can already get a qualitative
understanding of what happens from (4.3), (4.4). In fact,
setting $s_{0}=s(t=0)$, if the field equations are such that both
${1\over a}{da\over ds}{ds\over dt}$ and $Q$ remain finite
at $s_{0}$ (and similarly for $s_{f}=s(t=t_{f})$, where
$t_{f}$ is the time at which the torsion-free model reaches
the singularity in the future), a solution to (4.3), (4.4)
will exist for all values of $s$ and the model will be
nonspacelike geodesically complete. This qualitative
argument seems to suggest that, whatever the physical
source of torsion is (spin or theories with quadratic
Lagrangians etc.), nonspacelike geodesic completeness is a
concept of physical relevance even though test particles may
not move along geodesics.

An important comment is now in order. We have defined
geodesics exactly as one does
in general relativity (see [1], p. 33),
for reasons which will become even more
clear studying maximal timelike geodesics in sect. {\bf 5}.
However, our definition differs from the
one adopted in [10] (p. 1068).
In that paper, our geodesics are
just called autoparallel curves, whereas the authors interpret
as geodesics the curves of extremal length whose tangent
vector is parallelly transported according to the Christoffel connection.

Now, in view of the fact that the definition of timelike, null and
spacelike vectors is not affected by the presence of torsion, the whole
theory of causal structure [1] remains unchanged.
Combining this remark (also made in [10]) with the qualitative
argument concerning the geodesic equation, we here give the following
preliminary definition:
\vskip 0.3cm
{\it Definition} A $U_{4}$ space-time is singularity-free if it is
timelike and null
geodesically complete, where geodesics are defined as curves
whose tangent vector moves by parallel transport with respect to the full
$U_{4}$ connection.

This definition differs from the one given in [10] because we
rely on a different definition of geodesics, and it
has the drawbacks already illustrated in the beginning of
sect. {\bf 3}. However, it seems to have the following advantages:

1) it is a preliminary definition which
allows a direct comparison with
the corresponding situation in general relativity;

2) it is generic in that it does not depend on the specific physical
theory which is the source of torsion;

3) it has physical relevance as we have shown before looking at
a closed FRW model and at the causal structure [10].

The meaning of the remark 1) is that one can now try to make the same
(and eventually additional)
assumptions which lead to singularity theorems in general relativity, and
check whether one gets timelike and/or null geodesic incompleteness.
Indeed, the extrinsic curvature tensor and the vorticity which appears in
the Raychaudhuri equation will now explicitly contain the effects of
torsion, and it is not {\it a priori} clear what is going
to happen. Namely, if one adopts the above definition as a preliminary
definition of singularities in a $U_{4}$ space-time, the main
unsolved issues seem to be:

1) How can we explain from first principles that a space-time which is
nonspacelike geodesically incomplete may become nonspacelike geodesically
complete in the presence of torsion? And is the converse possible?

2) What happens in a $U_{4}$ space-time [10] under the
assumptions which lead
to the theorems of Penrose, Hawking and Geroch ?

Question 1) should not seem trivial in view of the FRW example
discussed before. In fact one should study the singularity
problem in a generic space-time. This is why we shall partially
study question 2) in the next section.
\vskip 10cm
\noindent
{\bf 5. - A singularity theorem without causality assumptions
for $U_{4}$ space-times.}
\vskip 1cm
In this section we shall denote by $R(X,Y)$ the four-dimensional
Ricci tensor with scalar curvature $R$, and by $K(X,Y)$ the extrinsic
curvature tensor of a spacelike three-surface. The energy-momentum tensor
will be written as $T(X,Y)$, so that the Einstein equations are
$$
R(X,Y)-{1\over 2}g(X,Y)R=T(X,Y) \; .
\eqno (5.1)
$$
In so doing, we are absorbing the $8\pi G$ factor into the definition
of $T(X,Y)$. A linear torsion-free connection will be denoted
by $\nabla$, so as to avoid confusion with ${\widetilde \nabla}$
appearing in (4.1).
For the case of general relativity, it was proved in [5] that
singularities must occur under certain assumptions, even though no
causality requirements are made. In fact, Hawking's result [1, 5]
states that space-time cannot be timelike geodesically complete if

1) $R(X,X)\geq 0$ for any nonspacelike vector $X$ (which can also be
written in the form: $T(X,X)\geq g(X,X){T\over 2}$),

2) there exists a compact spacelike three-surface $\Sigma$ without edge,

3) the trace $K$ of the extrinsic curvature tensor $K(X,Y)$ of $\Sigma$
is either everywhere positive or everywhere negative.

We are now going to study the following problem: is there a suitable
generalization of this theorem in the case of a $U_{4}$ space-time?
Indeed, a careful examination of Hawking's proof (see [1],
p. 273) shows that the arguments which are to be modified
in a $U_{4}$
space-time are the ones involving the Raychaudhuri equation and the results
which prove the existence or the nonexistence of conjugate points. We
are now going to examine them in detail.
\vskip 0.3cm
\leftline {\it I) Raychaudhuri equation.}
\vskip 0.3cm
The generalized Raychaudhuri equation in the ECSK theory of gravity has
been derived in [21, 22] (see also [23, 24]).
It turns out that, denoting by ${\widetilde \omega}_{ab}$
and $\sigma_{ab}$, respectively, the vorticity and the shear tensors, the
expansion $\theta$ for a timelike congruence of curves obeys the
equation
$$
{d\theta \over ds}=-(R(U,U)+2\sigma^{2}-2{\widetilde \omega}^{2})
-{\theta^{2}\over 3}
+{\widetilde \nabla}_{a}{\left(\dot U \right)}^{a} \; .
\eqno (5.2)
$$
In (5.2), $U$ is the unit timelike tangent vector, and we have set
$$
2\sigma^{2}\equiv \sigma_{ab}\sigma^{ab} \; , \; \; \; \;
2{\widetilde \omega}^{2}\equiv
\left(\omega_{ab}+{1\over 2}{\widetilde S}_{ab}\right)
\left(\omega^{ab}+{1\over 2}{\widetilde S}^{ab}\right) \; ,
\eqno (5.3)
$$
where $\omega_{ab}$ is the vorticity tensor
for the torsion-free connection $\nabla$,
and ${\widetilde S}_{bc}$ is obtained from the spin tensor
$\sigma_{\; \; bc}^{a}$ through a relation usually assumed to be of the
form [22, 25]
$$
\sigma_{\; \; bc}^{a}={\widetilde S}_{bc}U^{a} \; .
\eqno (5.4)
$$
\vskip 0.3cm
\leftline {\it II) Existence of conjugate points.}
\vskip 0.3cm
Conjugate points are defined as in general relativity [1],
but bearing in mind that now the Riemann tensor
is the one obtained from the connection ${\widetilde \nabla}$ appearing
in (4.1):
$$
R(X,Y)Z= {\widetilde \nabla}_{X}{\widetilde \nabla}_{Y}Z
-{\widetilde \nabla}_{Y}{\widetilde \nabla}_{X}Z
-{\widetilde \nabla}_{[X,Y]}Z \; .
\eqno (5.5)
$$
In general relativity, if one assumes that at $s_{0}$ one has
$\theta(s_{0})=\theta_{0}<0$, and $R(U,U)\geq 0$, everywhere, then one can
prove there is a point conjugate to $q$ along $\gamma(s)$ between
$\gamma(s_{0})$ and $\gamma \left(s_{0}-{3\over \theta_{0}}\right)$, provided
$\gamma(s)$ can be extended to $\gamma \left(s_{0}-{3\over
\theta_{0}}\right)$. This result is then extended to prove the existence
of points conjugate to a three-surface $\Sigma$ along $\gamma(s)$ within
a distance ${3\over \theta'}$ from $\Sigma$, where $\theta'$ is the initial
value of $\theta$ given by the trace $K$ of $K(X,Y)$, provided
$K<0$ and $\gamma(s)$ can be extended to that distance (see propositions
4.4.1 and 4.4.3 in [1]). This is achieved
studying an equation of the kind (5.2) where
${\widetilde \omega}^{2}=\omega^{2}$ is vanishing because $\omega_{ab}$
is constant and initially vanishing and the last term on the right-hand
side vanishes as well. However, in the ECSK theory,
${\widetilde \omega}^{2}$ will still contribute in view of (5.3).
Thus the inequality
$$
{d\theta \over ds}\leq -{\theta^{2}\over 3}
$$
can only make sense if we assume that
$$
R(U,U)-2{\widetilde \omega}^{2}\geq 0 \; ,
\eqno (5.6)
$$
where we do not strictly need to include $2\sigma^{2}$ on the left-hand side
of (5.6) because $\sigma^{2}$ is positive [1, 24].
If (5.6) holds true, we can write (see (5.2) and set there
${\widetilde \nabla}_{a}{\left(\dot U \right)}^{a}=0$)
$$
\int_{\theta_{0}}^{\theta}y^{-2}dy \leq -{1\over 3}\int_{s_{0}}^{s}dx
\; ,
\eqno (5.7)
$$
which implies
$$
\theta \leq {3\over {s-\left(s_{0}-{3\over \theta_{0}}\right)}}
\; ,
\eqno (5.8)
$$
where $\theta_{0}<0$. Thus $\theta$ becomes infinite and there are
conjugate points for some $s\in \left]s_{0},s_{0}-{3\over \theta_{0}}\right]$.
However, (5.6) is a restriction on the torsion tensor. In fact, the
equations of the ECSK theory are given by (5.1) plus another one more
suitably written in component language in the form [25]
$$
S_{\; \; bc}^{a}-\delta_{b}^{a}S_{\; \; dc}^{d}
-\delta_{c}^{a}S_{\; \; bd}^{d}
=\sigma_{\; \; bc}^{a} \; .
\eqno (5.9)
$$
In following [25] we temporarily choose (up to (5.12)) a
convention opposite to the one of sec. {\bf 4}, working with a
torsion tensor $S_{\; \; bc}^{a}=-S_{\; \; cb}^{a}$,
rather than $S_{bc}^{\; \; \; a}=-S_{cb}^{\; \; \; a}$.

In (5.9)
we have absorbed the $8\pi G$ factor into the definition of
$\sigma_{\; \; bc}^{a}$, whereas this is not done in (5.4). Setting
$\epsilon=g(U,U)=-1$, $\rho=8\pi G$, the insertion of
(5.4) into (5.9) and the multiplication by $U_{a}$ yields
$$
{\widetilde S}_{bc}={1\over \rho \epsilon}
\left(U_{a}S_{\; \; bc}^{a}-U_{b}S_{\; \; dc}^{d}
-U_{c}S_{\; \; bd}^{d} \right) \; ,
\eqno (5.10)
$$
which implies, defining
$$ \eqalignno{
f(\omega,\omega S) &\equiv \omega_{ab}\omega^{ab}+{1\over 2}
\omega_{ab}{\widetilde S}^{ab}+{1\over 2}{\widetilde S}_{ab}
\omega^{ab}\cr
&=\omega_{ab}\omega^{ab}+{\omega_{ab}\over 2\rho \epsilon}
\left(U_{f}S^{fab}-U^{a}S_{\; \; h}^{h \; \; b}
-U^{b}S_{\; \; \; h}^{ha}\right)\cr
&+{\omega^{ab}\over 2\rho \epsilon}
\left(U_{f}S_{\; \; ab}^{f}
-U_{a}S_{\; \; hb}^{h}-U_{b}S_{\; \; ah}^{h}\right)
&(5.11)\cr}
$$
and using (5.3) and (5.6), that
$$ \eqalignno{
8{\widetilde \omega}^{2}&=4f(\omega,\omega S)
+g^{bl}g^{cm}{\widetilde S}_{bc}{\widetilde S}_{lm}\cr
&=4f(\omega,\omega S)+
\rho^{-2} \Bigr(U_{a}S^{alm}-U^{l}S_{\; \; d}^{d \; \; \; m}
-U^{m}S_{\; \; \; d}^{dl}\Bigr)
\Bigr(U_{f}S_{\; \; lm}^{f}-U_{l}S_{\; \; fm}^{f}
-U_{m}S_{\; \; lf}^{f}\Bigr)\cr
&\leq 4 R(U,U) \; .
&(5.12)\cr}
$$
Indeed some cases have been studied (see for example [21])
where $\omega_{ab}$ is vanishing.
However, we here prefer to write the equations in general form.
Moreover, in extending (5.8) so as to prove the existence of conjugate
points to spacelike three-surfaces, the assumption $K<0$ on the trace
$K$ of $K(X,Y)$ also implies another condition on the torsion tensor. In
fact, denoting by $\chi(X,Y)$ the tensor obtained from the metric and from
the lapse and shift functions as the extrinsic curvature in general
relativity, in a $U_{4}$ space-time one has
$$
K(X,Y)=\chi(X,Y)+\lambda(X,Y) \; ,
\eqno (5.13)
$$
where the symmetric part of $\lambda(X,Y)$ (the only one which contributes
to $K$) is given by
$$
\lambda_{(ab)}=-2n^{\mu}S_{(a\mu b)} \; .
\eqno (5.14)
$$
In (5.14) we have changed sign with respect to [26] because that
convention for $K(X,Y)$ is opposite to Hawking's convention, and we are
here following Hawking so as to avoid confusion in comparing theorems.
Thus the condition $K<0$ implies the following restriction on torsion:
$$
\lambda=-2g^{ab}n^{\mu}S_{(a\mu b)}<-\chi \; .
\eqno (5.15)
$$
When (5.6) and (5.15) hold true, one follows exactly the same technique
which leads to (5.8) in proving there are points conjugate to a
spacelike three-surface.
\vskip 0.3cm
\leftline {\it III) Maximal timelike geodesics.}
\vskip 0.3cm
In general relativity, it is known (proposition 4.5.8 in [1])
that a timelike geodesic curve $\gamma$ from $q$ to $p$ is
maximal, if and only if there is no point conjugate to $q$ along
$\gamma$ in $(q,p)$. At the risk of boring the expert reader, we are now
going to sum up how this result is proved and then extended so as to rule out
the existence of points conjugate to three-surfaces. This last step will then
be enlightening in understanding what changes in a $U_{4}$ space-time.

We shall here follow the conventions of
subsect. {\bf 4}.5 of [1],
denoting by $L(Z_{1},Z_{2})$ the second derivative of the arc length
defined in (2.1), by $V$ the unit tangent vector ${\partial \over
\partial s}$ and by $T_{\gamma}$ the vector space consisting of all
continuous, piecewise $C^2$ vector fields along the timelike geodesic
$\gamma$ orthogonal to $V$ and vanishing at $q$ and $p$. We are here just
interested in proving that, if the timelike geodesic $\gamma$ from $q$
to $p$ is maximal, this implies there is no point conjugate to $q$. The
idea is to suppose for absurd that $\gamma$ is maximal but there is a
point conjugate to $q$. One then finds that $L(Z,Z)>0$, which in turn
implies that $\gamma$ is not maximal, against the hypothesis. This is achieved
taking a Jacobi field $W$ along $\gamma$ vanishing at $q$ and $r$, and
extending it to $p$ putting $W=0$ in the interval $[r,p]$. Moreover, one
considers a vector $M \in T_{\gamma}$ so that $g\left(M,{D\over \partial s}W
\right)=-1$ at $r$. In what follows, we shall just say that $M$ is suitably
chosen, in a way which will become clear later. One then defines
$$
Z \equiv \epsilon M +\epsilon^{-1}W \; ,
\eqno (5.16)
$$
where $\epsilon$ is positive and constant. Thus, the general formula for
$L(Z_{1},Z_{2})$
implies that (see lemma 4.5.6 of [1])
$$
L(Z,Z)=\epsilon^{2}L(M,M)+2L(W,M)+\epsilon^{-2}L(W,W)
=\epsilon^{2}L(M,M)+2 \; ,
\eqno (5.17)
$$
which implies that $L(Z,Z)$ is $>0$ if $\epsilon$ is suitably small, as
we anticipated. The same method is also used in proving there cannot be
points conjugate to a three-surface $\Sigma$ if the timelike geodesic
$\gamma$ from $\Sigma$ to $p$ is maximal. However, as proved in lemma
4.5.7 of [1], in the case of a three-surface $\Sigma$,
the formula for $L(Z_{1},Z_{2})$ is of the kind
$$
L(Z_{1},Z_{2})=F(Z_{1},Z_{2})-\chi(Z_{1},Z_{2}) \; ,
\eqno (5.18)
$$
where $\chi(X,Y)$ is the extrinsic curvature tensor of $\Sigma$. But we
know that in a $U_{4}$ space-time $\chi(X,Y)$ gets replaced by the
nonsymmetric tensor $K(X,Y)$ defined in
(5.13), (5.14), which can be
completed with the relation for the antisymmetric part of $\lambda(X,Y)$:
$$
\lambda_{[ab]}=-n^{\mu}S_{ba \mu} \; .
\eqno (5.19)
$$
Thus in $U_{4}$ theory
the splitting (5.16) leads to a formula of the kind (5.17)
where the requirement
$$
L(W,M)+L(M,W)=c>0
\eqno (5.20)
$$
will involve torsion because (5.18) gets replaced by
$$
L(Z_{1},Z_{2})={\widetilde F}(Z_{1},Z_{2})-K(Z_{1},Z_{2}) \; .
\eqno (5.21)
$$
Namely, the left-hand side of (5.20) will contain $K(W,M)+K(M,W)$.
Condition (5.20) also clarifies how to suitably choose $M$ in a
$U_{4}$ space-time. It is worth emphasizing that only $\lambda_{(ab)}$
contributes to (5.20) because the contributions of $\lambda_{[ab]}$
coming from $K(M,W)$ and $K(W,M)$ add up to zero. In proving (5.21),
the first step is the generalization of lemma 4.5.4 of [1]
to a $U_{4}$ space-time. This is achieved remarking that the
relation
$$
{\partial \over \partial u}g \left({\partial \over \partial t},
{\partial \over \partial t} \right)=
2g \left({D\over \partial u}{\partial \over \partial t},
{\partial \over \partial t}\right)
\eqno (5.22)
$$
is also valid in a $U_{4}$ space-time, where now ${D\over \partial u}$
denotes the covariant derivative along the curve with respect to the full
$U_{4}$ connection. In fact, denoting by $X$ the vector ${\partial \over
\partial t}$ and using the definition of covariant derivative along a curve
one finds
$$
{\partial \over \partial u}g(X,X)=2g \left({D\over \partial u}X,X \right)
+X^{a}X^{b}{D\over \partial u}g_{ab} \; ,
\eqno (5.23)
$$
where ${D\over \partial u}g_{ab}$ is vanishing if the connection obeys the
metricity condition, which is also assumed in a $U_{4}$ space-time
(see sect. {\bf 4} and [8]). In other words, the key role is
played by the connection which obeys the metricity condition, and
${\partial \over \partial u}g(X,X)$ will implicitly contain the effects of
torsion because of the relation
$$
{DX^{a}\over \partial u}\equiv {\partial X^{a}\over \partial u}
+\Gamma_{bc}^{\; \; \; a}{dx^{b}\over du}X^{c} \; .
\eqno (5.24)
$$
Although this point seems to be elementary, it plays a vital role in
leading to (5.21). This is why we chose to greatly emphasize it.

If we now compare the results discussed or proved in
I)-III) with p.
273 of [1], we are led to state the following singularity theorem:
\vskip 0.3cm
{\it Theorem.} The $U_{4}$ space-time of the ECSK theory cannot be
timelike geodesically complete if

1) $R(U,U)-2{\widetilde \omega}^{2} \geq 0$ for any
nonspacelike vector $U$;

2) $L(W,M)+L(M,W)=c>0$ as defined in (5.20), (5.21) and before;

3) there exists a compact spacelike three-surface $S$ without edge ;

4) the trace $K$ of the extrinsic curvature tensor $K(X,Y)$ of $S$ is
either everywhere positive or everywhere negative.

Conditions 1), 2) and 4) will then involve the torsion tensor
defined through (4.1). Indeed, condition 2)
can be seen as a
prerequisite, but we have chosen to insert it into the statement of the
theorem so as to present together all conditions which involve the
extrinsic curvature tensor $K(X,Y)$.
The compatibility of 1) with the field
equations of the ECSK theory is expressed by (5.12)
whenever (5.4) makes sense. Otherwise, (5.12) is to be replaced by a
different relation.
It is worth emphasizing that if we switch off torsion, condition 1)
becomes the one required in general relativity because, as
explained at p. 96-97 of [1], the vorticity of the
torsion-free connection vanishes wherever a
$3 \times 3$ matrix which appears in the
Jacobi fields is nonsingular. Finally, if ${\widetilde \nabla}_{a}
{\left(\dot U \right)}^{a}$ is not vanishing as we assumed so far (see
(5.2) and comment before (5.7)) following [21, 22],
condition 1) of our theorem is to be replaced by

($1'$) $R(U,U)-2{\widetilde \omega}^{2}-{\widetilde \nabla}_{a}
{\left(\dot U \right)}^{a} \geq 0$ for any nonspacelike vector $U$.
\vskip 1cm
\leftline {\bf 6. - Concluding remarks.}
\vskip 1cm
At first we have
taken the point of view according to which nonspacelike
geodesic incompleteness can be used as a preliminary definition of
singularities also in space-times with torsion. We have finally been able
to show under which conditions Hawking's singularity theorem without
causality assumptions can be extended to the space-time of the ECSK theory.
However, when we assume (5.4) and
we require consistency of the additional condition (5.6)
with the equations of the ECSK theory, we end up with the relation (5.12)
which explicitly involves the torsion tensor on the left-hand side
(of course, the torsion tensor is also present in $R(U,U)$ through the
connection coefficients, but this is an implicit appearance of torsion,
and it is better not to make this splitting). Also the conditions (5.15)
and (5.20) involve the torsion tensor in an explicit way
if one uses the formula (5.13).
This is why we interpret our result as an
indication of the fact that the presence of singularities in the ECSK theory
is less generic than in general relativity.
Our result is to be compared with [10]. In increasing order of
importance, the
differences between our work and their work are:

1) They look at the singularity theorem of Hawking and
Penrose in the ECSK theory, whereas we look at the singularity
theorem without causality assumptions in the ECSK theory.

2) We rely on a different definition of geodesics, as explained in
sect. {\bf 4}.

3) We emphasize the role played by the full extrinsic curvature tensor
and by the variation formulae in
$U_{4}$ theory, a remark which is absent in [10].

4) We keep the field equations of the ECSK theory in their original form,
whereas the authors in [10] cast them in a form analogous to general
relativity, but with a modified energy-momentum tensor which contains
torsion. We think this technique is not strictly needed. Moreover,
from a Hamiltonian point of view, the splitting of the Riemann tensor into
the one obtained from the Christoffel symbols plus the one explicitly
related to torsion does not seem to be in agreement with the choice of the
full connection as a canonical variable. In fact, if we look
for example at models with quadratic Lagrangians in $U_{4}$ theory, the frame
and the full connection are to be regarded as independent variables
(see [20]), and this choice of canonical variables has also been
made for the ECSK theory [27-29].

Problems to be studied for further research are the generalization to
$U_{4}$ space-times of the other singularity theorems in [1]
using our approach, and
of the results in [15] and [17] that we outlined in sect. {\bf 3}.
\vskip 1cm
\centerline {$* \; * \; *$}
\vskip 1cm
We here wish to express our deep gratitude to K. Wheeler and
P. D'Eath for invaluable moral support in undertaking the task
of writing this paper. We are also very much indebted to
P. Scudellaro and C. Stornaiolo for enlightening conversations
on theories with torsion, and for bringing to our attention
the references of sect. {\bf 5} and ref. [2]. Finally,
discussions with J. Stewart in the latest stages of our work
are gratefully acknowledged.
\vskip 1cm
\leftline {\it REFERENCES}
\vskip 1cm
\item {[1]}
S. W. HAWKING and G. F. R. ELLIS: {\it The Large-Scale
Structure of Space-Time} (Cambridge University Press,
Cambridge, 1973).
\item {[2]}
H. FUCHS, V. KASPER, D.-E. LIEBSCHER, V. MULLER and
H.-J. SCHMIDT: {\it Fortschr. Phys.}, {\bf 36},
427 (1988).
\item {[3]}
S. W. HAWKING: {\it Proc. R. Soc. London, Ser. A},
{\bf 294}, 511 (1966).
\item {[4]}
S. W. HAWKING: {\it Proc. R. Soc. London, Ser. A},
{\bf 295}, 490 (1966).
\item {[5]}
S. W. HAWKING: {\it Proc. R. Soc. London, Ser. A},
{\bf 300}, 187 (1967).
\item {[6]}
S. W. HAWKING and R. PENROSE: {\it Proc. R. Soc. London, Ser. A},
{\bf 314}, 529 (1970).
\item {[7]}
R. P. GEROCH: {\it Phys. Rev. Lett.}, {\bf 17}, 445 (1966).
\item {[8]}
F. W. HEHL, P. von der HEYDE, G. D. KERLICK and
J. M. NESTER: {\it Rev. Mod. Phys.}, {\bf 48}, 393 (1976).
\item {[9]}
J. M. NESTER: in {\it An Introduction to Kaluza-Klein Theories},
edited by H. C. LEE (World Scientific, Singapore, 1983), p. 83.
\item {[10]}
F. W. HEHL, P. von der HEYDE and G. D. KERLICK:
{\it Phys. Rev. D}, {\bf 10}, 1066 (1974).
\item {[11]}
J. K. BEEM and P. E. EHRLICH: {\it Global Lorentzian Geometry}
(Dekker, New York, N.Y., 1981).
\item {[12]}
S. GALLOT, D. HULIN and J. LAFONTAINE: {\it Riemannian Geometry}
(Springer-Verlag, Berlin, 1987).
\item {[13]}
K. GROVE: in {\it Differential Geometry}, edited by V. L. HANSEN
(Springer-Verlag, Berlin, 1987), p. 171.
\item {[14]}
B. O'NEILL: {\it Semi-Riemannian Geometry} (Academic Press,
New York, N.Y., 1983).
\item {[15]}
B. G. SCHMIDT: {\it Gen. Rel. Grav.}, {\bf 1}, 269 (1971).
\item {[16]}
R. P. GEROCH: {\it Ann. Phys.}, {\bf 48}, 526 (1968).
\item {[17]}
J. GRUSZCZAK, M. HELLER and Z. POGODA: {\it A singular
boundary of the closed Friedmann Universe}, Cracow
preprint, TPJU 11 (1989).
\item {[18]}
B. BOSSHARD: {\it Commun. Math. Phys.}, {\bf 46},
263 (1976).
\item {[19]}
R. A. JOHNSON: {\it J. Math. Phys. (N.Y.)}, {\bf 18},
898 (1977).
\item {[20]}
G. ESPOSITO: {\it Nuovo Cimento B}, {\bf 104}, 199 (1989).
\item {[21]}
J. M. STEWART and P. H\'AJICEK: {\it Nat. Phys. Sci.},
{\bf 244}, 96 (1973).
\item {[22]}
J. TAFEL: {\it Phys. Lett. A}, {\bf 45}, 341 (1973).
\item {[23]}
A. K. RAYCHAUDHURI: {\it Phys. Rev. D}, {\bf 12}, 952 (1975).
\item {[24]}
A. K. RAYCHAUDHURI: {\it Theoretical Cosmology}
(Clarendon Press, Oxford, 1979).
\item {[25]}
M. DEMIANSKI, R. DE RITIS, G. PLATANIA, P. SCUDELLARO
and C. STORNAIOLO: {\it Phys. Rev. D}, {\bf 35},
1181 (1987).
\item {[26]}
M. PILATI: {\it Nucl. Phys. B}, {\bf 132}, 138 (1978).
\item {[27]}
J. ISENBERG and J. NESTER: in {\it General Relativity
and Gravitation}, edited by A. HELD (Plenum Press,
New York, N.Y., 1980), p. 23.
\item {[28]}
L. CASTELLANI, P. VAN NIEUWENHUIZEN and M. PILATI:
{\it Phys. Rev. D}, {\bf 26}, 352 (1982).
\item {[29]}
R. DI STEFANO and R. T. RAUCH: {\it Phys. Rev. D},
{\bf 26}, 1242 (1982).

\bye